\documentclass[%
 reprint,
 superscriptaddress,
 twocolumn,
 amsmath,amssymb,
 aps,
prx,
floatfix,
]{revtex4-2}
\usepackage{graphicx}% Include figure files
\usepackage{dcolumn}% Align table columns on decimal point
\usepackage{bm}% bold math
\usepackage{siunitx}
\usepackage{orcidlink}

\usepackage{import}

\usepackage[columnwise]{lineno}
\usepackage[normalem]{ulem}

\usepackage{enumitem}

\usepackage{hyperref}% add hypertext capabilities

\begin{document}
%%%%%%%%%%%%%%%%%%%%%%%%%%%%%%%%%%%%%%%%%%%%
\title{Report on reproducibility in condensed matter physics}% Force line breaks with \\

\author{A. Akrap}
\affiliation{Department of Physics, Faculty of Science, University of Zagreb, HR-10000 Zagreb, Croatia}

\author{D. Bordelon}
\affiliation{Pittsburgh Supercomputing Center, Pittsburgh, PA 15213, USA}

\author{S. Chatterjee}
\affiliation{Department of Physics, Carnegie Mellon University, Pittsburgh, PA 15213, USA}

\author{E. D. Dahlberg}
\affiliation{School of Physics and Astronomy, The University of Minnesota, Minneapolis, MN 55455, USA}

\author{R. P. Devaty}
\affiliation{Department of Physics and Astronomy, University of Pittsburgh, Pittsburgh, PA 15260, USA}

\author{S. M. Frolov}
\email{frolovsm@pitt.edu}
\affiliation{Department of Physics and Astronomy, University of Pittsburgh, Pittsburgh, PA 15260, USA}

\author{C. Gould}
\affiliation{Faculty for Physics and Astronomy (EP3), Universitat Würzburg, Am Hubland, Würzburg, Germany}
\affiliation{Institute for Topological Insulators, Am Hubland, Würzburg, Germany}

\author{L. H. Greene}
\affiliation{National High Magnetic Field Laboratory, Florida State University, FL 32310, USA}
\affiliation{Department of Physics, Florida State University, FL 32306, USA}

\author{S. Guchhait}
\affiliation{Department of Physics and Astronomy, Howard University, Washington, DC 20059, USA}

\author{J. J. Hamlin}
\affiliation{Department of Physics, University of Florida, Gainesville, FL 32611, USA}

\author{B. M. Hunt}
\affiliation{Department of Physics, Carnegie Mellon University, Pittsburgh, PA 15213, USA}

\author{M. J. A. Jardine}
\affiliation{Department of Materials Science and Engineering, Carnegie Mellon University, Pittsburgh, PA 15213, USA}

\author{M. Kayyalha}
\affiliation{Department of Electrical Engineering, The Pennsylvania State University, University Park, PA 16802, USA}
\affiliation{Materials Research Institute, The Pennsylvania State University, University Park, PA 16802, USA}

\author{R. C. Kurchin}
\affiliation{Department of Materials Science and Engineering, Carnegie Mellon University, Pittsburgh, PA 15213, USA}

\author{V. Kozii}
\affiliation{Department of Physics, Carnegie Mellon University, Pittsburgh, PA 15213, USA}

\author{H. F. Legg}
\affiliation{Department of Physics, University of Basel, 4056 Basel, Switzerland}
\affiliation{SUPA, School of Physics and Astronomy, University of St. Andrews, St Andrews KY16 9SS, UK}

\author{I. I. Mazin}
\affiliation{Department of Physics and Astronomy, George Mason University, Fairfax, VA 22030, USA}
\affiliation{Center for Quantum Science and Engineering, George Mason University, Fairfax, VA 22030, USA}

\author{V. Mourik}
\affiliation{JARA-FIT Institute for Quantum Information, Peter Grünberg Institute (PGI-11), Forschungszentrum Jülich GmbH, 52074 Aachen, Germany}

\author{A. B. Özgüler}
\affiliation{Department of Mechanical Engineering and Materials Science, University of Pittsburgh, Pittsburgh, PA 15261, USA}
\affiliation{Haas School of Business, University of California, Berkeley, CA 94720, USA}

\author{J.~Peñuela-Parra}
\affiliation{Department of Mechanical and Materials Science, University of Pittsburgh, Pittsburgh, PA 15261, USA}

\author{B. Seradjeh}
\affiliation{Department of Physics, Indiana University, Bloomington, IN 47405, USA}
\affiliation{Indiana University Center for Spacetime Symmetries, Bloomington, IN 47405, USA}
\affiliation{Quantum Science and Engineering Center, Indiana University, Bloomington, IN 47405, USA}

\author{B. Skinner}
\affiliation{Department of Physics, Ohio State University, Columbus, OH 43210, USA}

\author{K. F. Quader}
\affiliation{Department of Physics, Kent State University, Kent, OH, 44242, USA}

\author{J.~P.~Zwolak}
\email{jpzwolak@nist.gov }
\affiliation{National Institute of Standards and Technology, Gaithersburg, MD 20899, USA}
\affiliation{Joint Center for Quantum Information and Computer Science, University of Maryland, College Park, MD 20742, USA}
\affiliation{Department of Physics, University of Maryland, College Park, MD 20742, USA}

\date{\today}
%%%%%%%%%%%%%%%%%%%%%%%%%%%%%%%%%%%%%%%%%%%%%%%%%%%%%%%%%%%%%
\begin{abstract}
We present recommendations to improve reproducibility and replicability in condensed matter physics. 
This area of physics has consistently produced both fundamental insights into the workings of matter and transformative inventions. 
Our recommendations result from a collaboration that includes researchers from academia and government laboratories, scientific journalists, legal professionals, representatives of publishers, professional societies, and other experts. 
The group met in person in May 2024 at a conference at the University of Pittsburgh to discuss the growing challenges related to research reproducibility and replicability in condensed matter physics. 
In this report, we discuss best practices and policies at all stages of the scientific process to safeguard the value of condensed matter.
We hope this report will lay the groundwork for a broader conversation to develop subfield-specific recommendations.
\end{abstract}

%\keywords{Suggested keywords}%Use showkeys class option if keyword
                              %display desired
\maketitle
%%%%%%%%%%%%%%%%%%%%%%%%%%%%%%%%%%%%%%%%%%%%%%%%%%%%%%%%%%%%%
%\pacs{}% insert suggested PACS numbers in braces on next line
%%%%%%%%%%%%%%%%%%%%%%%%%%%%%%%%%%%%%%%%%%%%%%%%%%%%%%%%%%%%%

%%%%%%%%%%%%%%%%%%%%%%%%%%%%%%%%%%%%%%%%%%%%%%%%%%%%%%%%%%%%%
% \tableofcontents{}
%%%%%%%%%%%%%%%%%%%%%%%%%%%%%%%%%%%%%%%%%%%%%%%%%%%%%%%%%%%%%
% \linenumbers
%%%%%%%%%%%%%%%%%%%%%%%%%%%%%%%%%%%%%%%%%%%%%%%%%%%%%%%%%%%%%
\section{Introduction}
%%%%%%%%%%%%%%%%%%%%%%%%%%%%%%%%%%%%%%%%%%%%%%%%%%%%%%%%%%%%%
Condensed matter physics (CMP) is a vast and consequential area of physics research. 
According to the American Physical Society (APS), the Division of Condensed Matter Physics, encompassing condensed matter and related areas, is by far the largest division, with nearly 6,700 members. 
Originally defined as the study of the physical properties of solid materials, such as metals and semiconductors, the field has expanded its focus in recent decades to include more exotic phases of matter, including magnets, superconductors, and, more recently, topological materials. 
Soft matter, materials science, and quantum information are all closely related fields. 
Advances in CMP have deepened our understanding of the matter that surrounds us, advanced fields such as particle physics and astrophysics, and enabled many of our most recognizable technological advances. 
Examples of inventions made possible by research in CMP include light-emitting diodes, integrated circuits, magnetic resonance imaging and nuclear magnetic resonance machines, lasers, cell phones, the global positioning system, hard drives, and more. 
There is no evidence that progress in our understanding of matter has slowed, which means future technological advances are likely.

Other impactful sciences, such as social and biomedical sciences, have recently been grappling with the replication crisis, in which a significant fraction of published results cannot be verified by other researchers~\cite{Munafo17, Errington21, Freese22}. 
While the causes of and solutions to this phenomenon are still not entirely clear, its roots are believed to involve biases in the presentation of results, accompanied by nontransparent methodology, which facilitates unverifiable claims. 
The writers of this report share a concern that a growing replication crisis exists in CMP that needs to be addressed.
As in other sciences, CMP shows a bias toward publishing and publicizing surprising or exciting claims. 
In CMP, a curiosity-driven field, there is also a bias toward novel rather than more mundane narratives (an effect named the “Inverse Occam’s Razor” by Mazin~\cite{Mazin22}). 
At the same time, the presentation of results in more than a few cases tends to lack sufficient detail to facilitate replication and verification. 
Reputation and authority serve as substitutes for which results to consider trustworthy.

A common belief is that physics, as a field, is immune to the replication crisis since its experiments are considered data-rich compared to, e.g., social or medical sciences. 
We often hear that in the natural sciences, any study can, in principle, be replicated and verified because the laws of nature are invariant. 
Unreliable work will eventually be weeded out if no one can repeat it (“science corrects itself”). 
However, in reality, it is not clear whether a particular direction fades because the community has lost faith in its central claims or rather because attention has shifted to a new topic. 
Furthermore, behind every disproven claim are months of work, precious funding resources that are typically not dedicated to replication studies, and diverted careers, often those of young researchers. 
As long as it is easier to publish sensational claims than to support efforts to replicate and verify them, enhancing general reproducibility and replicability will remain crucial.

%%%%%%%%%%%%%%%%%%%%%%%%%%%%%%%%%%%%%%%%%%%%%%%%%%%%%%%%%%%%%
\subsection*{What happened at the conference}
%%%%%%%%%%%%%%%%%%%%%%%%%%%%%%%%%%%%%%%%%%%%%
This report summarizes the conclusions and recommendations from the working groups at the \textit{International Conference on Reproducibility in Condensed Matter Physics}~\cite{ICRCMP}, which took place at the University of Pittsburgh from May 9 to May 11, 2024. 
This conference brought together junior and senior researchers, journal editors, government program managers and inspectors, journalists, and legal professionals to discuss reproducibility, replicability, bias, scientific integrity, and collaboration in CMP. 

Talks and panels considered recent case studies of reproducibility and replicability issues in the field, including high-pressure superconductivity, unconventional superconductivity, the study of Majorana zero modes, and density functional theory. 
Talks and panel discussions were recorded, and most recordings are available on YouTube~\cite{ICRCMP-rec}.

In addition to talks and panel discussions, the conference included significant time (7 hours over 3 days) for group discussions on reproducibility and replicability in the field of CMP and how they can be addressed. 
The conclusions of those discussions are summarized in this document, which we hope will serve as a resource for scientists, publishers, universities, research laboratories, and funding institutions as they develop policies and practices to mitigate reproducibility issues and foster a cultural change toward more reproducible and replicable research. 
Some contemporaneous news reports about the conference and discussion articles have already appeared in the press~\cite{Padavic-Callaghan24, Houle24, Chen24}.

The conference received support from the National Science Foundation (NSF) and the Pittsburgh Quantum Institute. 
A documentary film essay~\cite{ICRCMP-documentary} and several interviews with participants~\cite{ICRCMP-interviews} were produced with support from the Julian Schwinger Foundation. 
NSF representatives were present at the conference and participated in talks and panels but did not attend the working group meetings. 
Representatives of the publishers participated in the working groups but did not take part in the preparation of the report following the conference. 
A. Akrap was appointed as an Associate Editor of Physical Review B (PRB) while participating in the conference and writing this report. 
In this, she was representing herself as an active researcher in the field and not as a PRB Associate Editor.

%%%%%%%%%%%%%%%%%%%%%%%%%%%%%%%%%%%%%%%%%%%%%%%%%%%%%%%%%%%%%
\section{Key conclusions of the report}
%%%%%%%%%%%%%%%%%%%%%%%%%%%%%%%%%%%%%%%%%%%%%%%%%%%%%%%%%%%%%
\textit{Reproducible research offers benefits to researchers and the public.} 
Presenting research findings in a reproducible and replicable manner is critical for validating findings by fellow scientists (mitigating errors in interpretation, confirmation bias, and fraud), benchmarking progress by subsequent investigators, follow-on analyses, synergistic effects, and long-term preservation of work and results.

\textit{Sharing data and code is critical.} 
Increased sharing of primary research materials is necessary to reproduce studies. 
This practice should be implemented community-wide.  
Fortunately, the infrastructure for making data available exists in the form of free public repositories and is adequate for the vast majority of experimental and numerical studies in CMP. 
Nonetheless, additional technique-specific guidance on which material to share and how to share it would be valuable.

\textit{We need a culture of advocating for reproducibility and replicability.} 
In the longer term, making our field robust and replicable will require changes to how research is organized, recorded, and presented. 
Individual members of the community can shift the status quo by improving their own practices, requesting changes to procedures within our institutions, and advocating for changes to policy and its implementation. 
See the list of recommendations in this report for ideas.

\textit{Recommendations of the report: In brief}.
\noindent Best practices for condensed matter researchers:
\begin{enumerate}[label=(\arabic*), topsep=-3pt, itemsep=-3pt]
    \item When reporting experiments, include the full ranges of experimental parameters studied. 
    \item All authors of a manuscript should have access to all primary data and analysis files/software before they are asked to consent to publication. 
    \item The manuscript, its Appendixes, and Supplemental Material should contain enough information for replication.
    \item Authors should release as much primary data as possible and practical with a paper. 
    \item Strive to achieve the FAIR (Findable, Accessible, Interoperable, and Reusable) Data Standard~\cite{Wilkinson16}, but, if not possible, share materials in the form you have them. 
    \item A manuscript should include all reasonable presentations of the data.
    \item Authors should specify full data processing steps from original raw data to published figures. 
    \item A manuscript should summarize all plausible interpretations of results. 
    \item Information about the number of samples tested should be reported. 
    \item Research groups should have a stated policy and timeline for releasing in-house developed code.
    \item Subcommunities of CMP should develop subfield-specific minimum reporting standards.
    \item Authors should promptly retract or correct their own prior work if they discover errors post publication.
\end{enumerate}
\vspace{4pt}

\noindent Recommendations for scientific publishers:
\begin{enumerate}[label=(\arabic*), topsep=-3pt, itemsep=-3pt]
    \item Refocus on scientific validity rather than subjective appeal criteria for publication.
    \item Require availability of complete data and scripts in citable repositories.
    \item Formulate clear statements of acceptance criteria for replication efforts.
    \item Create review requirements that are specific to Supplemental Material.
    \item Develop specific guidelines for assessing the reproducibility and replicability of submissions.
    \item Consider adopting an open review process.
    \item Draft and publicize more transparent retraction and correction policies and procedures.
\end{enumerate}
\vspace{4pt}

\noindent Recommendations for universities and research laboratories:
\begin{enumerate}[label=(\arabic*), topsep=-3pt, itemsep=-3pt]
    \item Consider not only publications, but also datasets and code as research outputs.
    \item Recognize reproducibility and replicability practices in promotion cases. 
    \item De-emphasize publication indices in evaluating the productivity of individual scientists.
    \item Follow government-mandated procedures for investigating misconduct, and inquire into the full factual basis of each report.
    \item Develop a clear definition of unreliable science beyond the definition of misconduct.
    \item Actively protect whistleblowers.
\end{enumerate}
\vspace{4pt}

\noindent Recommendations for government and scientific funding organizations:
\begin{enumerate}[label=(\arabic*), topsep=-3pt, itemsep=-3pt]
    \item Consider funding work that explicitly aims to replicate prior results.
    \item Require grant proposals to address reproducibility and replicability in their planning.
    \item Evaluate and reward the reproducibility and replicability of research.
    \item Develop policies requiring grant recipient organizations to report allegations of scientific misconduct to the funding agency.
\end{enumerate}
\vspace{4pt}

%%%%%%%%%%%%%%%%%%%%%%%%%%%%%%%%%%%%%%%%%%%%%%%%%%%%%%%%%%%%%
\section{Background: Reproducibility and replicability in condensed matter physics}
%%%%%%%%%%%%%%%%%%%%%%%%%%%%%%%%%%%%%%%%%%%%%%%%%%%%%%%%%%%%%
The 2019 National Academies of Sciences, Engineering, and Medicine (NASEM) Report on “Reproducibility and Replicability in Science” offers a comprehensive summary of the consensus understanding of the problem~\cite{NASEM19}. 
While it does not explicitly discuss CMP, the report offers many relevant points. 

At the same time, a reader is likely to walk away from the report with the impression that replication issues largely do not concern physics. 
This is how the National Academies report talks about the difference between social and physical sciences: \textit{“[...] studies that are conducted in the relatively more controllable systems will replicate with greater frequency than those that are in less controllable systems [...].”} 
The example the report gives is a measurement of the semiconductor gap using tunneling microscopy.
Yet, physicists who perform these measurements consider the quantitative determination of the bandgap from tunneling spectra to be a challenging task, sensitive to the analysis performed, which may affect replicability~\cite{Feenstra09}.

The NASEM report discusses replicability vs. reproducibility, defining them as two distinct methods of verification. 
\textit{Reproducibility} is \textit{“obtaining consistent results using the same input data; computational steps, methods, and code; and conditions of analysis.”} 
In the context of a given paper, reproducibility may be understood as \textit{“the ability to re-create all of the figures and numbers [...] from the code and data provided by the authors”}~\cite{Leek17}. 
In contrast, \textit{replicability} is \textit{“obtaining consistent results across studies aimed at answering the same scientific question, each of which has obtained its own data”}~\cite{NASEM19}.
For a given paper, replicability is then \textit{“the ability to re-perform the experiments and computational analyses... and arrive at consistent results”}~\cite{Leek17}. 
A simple way to understand the distinction is that reproducibility concerns existing data, while replicability involves generating new data. 
From these definitions, it is clear that the degree of either replicability or reproducibility is directly dependent on the availability of the primary research materials (data, code, materials and samples, protocols) and contextual metadata (conditions in the laboratory, experimental device parameters, configuration of parameters not actively measured or changes in the experiment).

In communication about research verification, reproducibility, and replicability are often used interchangeably. 
However, the NASEM definitions are helpful and are followed as much as possible in this report. 
We provide their elaboration in the context of CMP and other fields that similarly involve largely exploratory research. 
For example, even when the figures of a paper can be reproduced using the original raw data they display, some of the data selection or processing steps may have been chosen by the authors in such a way that the conclusions of a study will not hold when additional data that were not included in the figure are considered. 

When it comes to replication, we often hear the term applied to cases in which two groups performed vastly different studies, e.g., using two very different techniques such as transport and neutron scattering. 
However, even if both studies were aimed at the same underlying physics questions, they would not constitute replications of each other, and their conclusions should be compared with caution. 
In some cases, when different studies arrive at consistent conclusions, this may be a result of a phenomenon known as “confirmation bias,” which has been documented in physics since at least Millikan’s oil drop experiment~\cite{Feynman74}.

There are other aspects of how the CMP field is organized that are important to consider when comparing it to other research areas. 
The majority of CMP studies are conducted by small teams of no more than 10 scientists, working either in the same laboratory or in collaboration among a few research groups. 
The subject matter is highly diverse, spanning sample types and techniques from optics to electronics, and from bulk crystals to nanoscale objects. 
Experimental apparatus and numerical codes are often bespoke. 
Samples that are nominally the same nonetheless differ in subtle details that can affect the outcomes. 
While CMP relies on a common set of foundational theories, new mathematical models are often developed for each study, complicating both meta-analysis and comparisons across studies. 
While some fields have responded to the replication crisis by implementing preregistration of hypotheses and study design, such preregistration may not be helpful in the context of exploratory research.  
CMP studies are often costly and time-consuming. 
These factors contribute to the challenge of assessing and improving reproducibility and replicability. 
Despite these differences, we argue that there are important steps that our community can take toward this goal.

The NASEM report understandably discusses statistical analysis of data, including uncertainty quantification, which is commonly used in other fields to identify and quantify replication issues. 
Counterintuitively, applying statistical techniques is often not feasible in CMP, because results are reported on a small number of samples, each providing a unique signature of an effect. 
For similar reasons, meta-studies are also complicated and rarely performed. 
This can be due to a complex and unique experimental apparatus or due to high computational complexity. 
However, a large factor is the lack of common standards for performing and reporting measurements. 
Due to these factors, some subfields of CMP, including mesoscopic physics and quantum materials, are in practice relatively qualitative.

%%%%%%%%%%%%%%%%%%%%%%%%%%%%%%%%%%%%%%%%%%%%%%%%%%%%%%%%%%%%%
\subsection{Conflicting incentives lead to nonreproducible research}
%%%%%%%%%%%%%%%%%%%%%%%%%%%%%%%%%%%%%%%%%%%%%
A large part of the challenge with reproducibility and replicability arises from several clashes of interests and incentives, which together prevent the establishment of a clear impetus toward accurate and comprehensive reporting of results. 
For example, while most researchers agree that results should be correct and are trained in the ways of preserving and analyzing data reproducibly, the professional rewards of a high-profile publication might sometimes outweigh the disincentives associated with sloppy, inaccurate, or even fraudulent work. 
This imbalance between individual researchers’ incentives and disincentives exists at all career levels, wherever there is pressure to claim significant advances and secure high-profile publications, which factor into decisions about grant funding, job offers, and tenure and promotion decisions. 

Another clash is evident in the fact that part of the value of a result can derive from how difficult it is for other research groups to replicate. 
Thus, scientists may have a vested interest in not making it easy for other groups to replicate the methods or code underlying their results, which also hinders reproducibility and replicability. 

Scientists and the public alike favor results that are perceived as striking and valuable. 
Publishers are generally interested in research of the highest quality, which implies that it is replicable.  
However, coupled with the trend to gamify publishing in high-impact venues, this preference creates conditions for positive bias in the reporting of outcomes. 
Publishers are often willing to push the envelope when it comes to manuscripts that are likely to boost a journal’s impact, and such papers are often published quickly at the risk of future scrutiny post-publication. 
In some high-profile cases, the claims are disproven within weeks of publication~\cite{Arute19, Kim23, Zhu23, King25, Pan22, Kalai23, Tindall24, Tindall25}. 
Anecdotally, many members of the CMP community perceive a historic shift in the narrative of papers from a neutral presentation of results to flashier stories. 
In recent decades, a declining proportion of published results are negative~\cite{Fanelli12}. 
Publication bias toward claims phrased as new and significant might skew editors’ and referees’ perceptions when evaluating manuscripts, shifting the focus to the novelty and noteworthiness of the results rather than their accuracy.   
Institutions also have a vested interest in allowing, and even encouraging, their affiliated authors to publish work that can generate publicity and secure grant funding. 
If administrators receive a complaint about the inaccurate presentation of results or potential misconduct, they may be disinclined to commission a thorough and objective investigation and may choose not to request corrections and retractions, or to publish findings of misconduct, in the belief that this will better protect the institution's reputation and, perhaps, their own.

Funding agencies have a clear interest in producing reliable research that is not a waste of tax dollars.
At the same time, program officers also want to maintain their own continued budgets, which can be revised annually. 
Perhaps for this reason, agency personnel sometimes promote and fund research perceived to have a high impact, unintentionally deprioritizing work with the potential to validate, build on, or, alternatively, discredit high-impact claims. 

Grant duration cycles that are short on the scale of a typical study create strong pressures on applicants to submit “transformative” proposals and grant reports, which, in turn, pressure them to report positive results as products of the grant. 
Funding bias can be more pernicious for reproducibility and replicability than publication bias, since it alters the priorities for work whose results have yet to be produced.

%%%%%%%%%%%%%%%%%%%%%%%%%%%%%%%%%%%%%%%%%%%%%%%%%%%%%%%%%%%%%
\subsection{How does nonreproducible research manifest itself in practice?}
%%%%%%%%%%%%%%%%%%%%%%%%%%%%%%%%%%%%%%%%%%%%%
While most researchers intuitively understand what a nonreplicable study is, it may be helpful to consider the types of nonreproducibility. 
As identified by NASEM, the sources of failure to replicate include inadequate record-keeping, nontransparent reporting, obsolescence of digital formats, mistakes or errors in the reproducer’s process, and barriers in the research culture~\cite {NASEM19}. 
Some specific examples include:
\begin{enumerate}[label=(\arabic*), topsep=-4pt, itemsep=-4pt]
    \item The data supporting the study are not online, nor available from the authors upon request. 
    In some cases, there is only limited data sharing. 
    The latter instances include decisions to release only part of the data or to release data in formats that are difficult to access or analyze (e.g., a PDF containing large volumes of tabular data, or JPG or other nonvector images of digital data).
    \item Protocols, other experimental information, and/or descriptive metadata that are necessary for replicating the setup and/or for interpreting data are not available. 
    Software used to process data is unavailable, or supporting libraries have evolved over time, perhaps breaking previously working code.
    \item Methods and original raw data are available somewhere, but were not shared in full due to the large file size, which makes storage/transfer prohibitive. 
    Relatedly, methods and original raw data are available, but computation for processing/analysis is prohibitive (perhaps requiring specialized, high-performance computing facilities).
    \item Fabricated and/or cherry-picked data, and/or misleading visualization (i.e., research misconduct). 
    \item Errors in interpretation or methodology, confirmation bias, and other biases that are undetectable due to insufficient sharing of data, methods, code, materials, etc.
\end{enumerate}

%%%%%%%%%%%%%%%%%%%%%%%%%%%%%%%%%%%%%%%%%%%%%%%%%%%%%%%%%%%%%
\subsection{How prevalent is nonreproducible research in condensed matter physics?}
%%%%%%%%%%%%%%%%%%%%%%%%%%%%%%%%%%%%%%%%%%%%%
Using the definitions in the NASEM report, the majority of research in CMP is not immediately reproducible or replicable because published works typically do not provide sufficient data, code, access to samples or materials, or detailed replication instructions. 
While this in itself does not mean the published claims cannot be reproduced and replicated with additional effort, the absence of established verification protocols, research record-sharing standards, or peer-review checklists makes reproducing CMP works a real challenge. 
While there is a growing trend to include more information and data in Supplemental Material, the major conclusions are typically based on the three to five figures in the main manuscript and their brief descriptions. 
It is still a common practice to make available only the digital files necessary to reproduce the main text figures. 
And even when more data supporting the claims is provided, it is usually not clear whether this is all the relevant data, and whether the shared dataset truly represents the months or years of work that led to the final publication, or if, along the way, some data that could alter or amend the conclusions were removed from analysis.

Many essential parts of the quality control process in science are confidential, meaning that concerns raised with journals or institutions are not shared with the broader community. 
In practice, many concerns are never expressed at all due to a variety of factors, including an unsupportive culture, fear of retaliation, time-consuming or arcane procedures, or the absence of appropriate avenues altogether for doing so. 
As a result, it is difficult to assess the prevalence of unreliable research in a given field. 
Anecdotally, many community members are aware of at least one paper that they have no confidence in. 
Community surveys highlight that researchers have serious concerns~\cite{Baker16, Houle23}. 
For example, 12~\% of junior APS members report witnessing “less than truthful description of research techniques,” and 7~\% report observing data falsification. 
Informal estimates provided during the conference ranged from significantly lower to significantly higher percentages. 
Even if the lowest estimates are closer to the truth, given the field's volume, there is still significant research output that is flawed. 

As evidence of a replication crisis, several high-impact claims and entire research directions have been questioned in the past few years, both experimentally and computationally. 
On the theoretical side, results in fields like many-body localization and AI-powered materials prediction are beginning to spark debates about computational replicability~\cite{Sels23, Suntajs20, Leeman24}. 
Experimental examples of replication problems include the observation of a supersolid helium and room temperature superconductivity~\cite{Zhu23, Kim04, Kim12, Tamir19, Frolov23}.

In the most severe cases, several experimental works have been retracted from high-impact journals, such as \textit{Nature} and \textit{Science}, on high-priority topics including room-temperature superconductivity and Majorana fermions~\cite{Gazibegovic22, Snider22, Thorp22, Zhang21, Dasenbrock-Gammon23, Snider24}. 
Historically, one of the most notable examples of scientific fraud, the Bell Labs Schön affair~\cite{Reich09}, also originated from CMP. 
Given how challenging it is to correct or retract problematic works within the current system, these examples are among the most dramatic and may represent just the tip of the iceberg. 

There are signs that the community is aware of these problems. 
Several opinion articles were published, questioning the methodology used in the field and expressing concerns about the state of play~\cite{Mazin22, Frolov21, Zelezny23, Zwolak24}. 
Efforts to reproduce some of the claims in the field have been made. 
In a number of these cases, conclusions did not replicate or alternative explanations have been put forward~\cite{Yu21, Kayyalha20, Valentini21, Valentini22, Tamir19, Dartiailh21, DeCecco16, Zhang22, Hlawenka18, Santos-Cottin23, Faugeras10, Lefrancois22, Zhu23}. 
The more traditional form of debating research results, in the form of comments on published works, often touches on replication issues. 
There is hope that efforts to enhance replicability can be expanded in the future through changes at the cultural, procedural, and policy levels.

%%%%%%%%%%%%%%%%%%%%%%%%%%%%%%%%%%%%%%%%%%%%%%%%%%%%%%%%%%%%%
\section{Open Science policies}
%%%%%%%%%%%%%%%%%%%%%%%%%%%%%%%%%%%%%%%%%%%%%%%%%%%%%%%%%%%%%
While national, institutional, and some corporate policies have long demanded the preservation of research materials and their availability on demand, there is a new worldwide push toward what is called “open science”. 
This is understood as more transparent reporting of research outcomes, which is universally perceived as beneficial for reproducibility and replicability.
While not all policies are finalized, and enforcement of new policies is expected to lag behind their introduction, a proactive approach will benefit our community. 
CMP researchers should prepare for these shifts. 
We provide ideas for what this may look like in the next section.

The transition toward open science is largely driven by the digitalization of research records and by the advance of online storage platforms. 
At the technological level, solutions for data sharing are already available, except for larger volumes of primary data, such as those exceeding 1 TB. 
For gigabyte-scale datasets, there are robust repositories (Zenodo, Harvard Dataverse, Open Science Framework, Figshare, Hepdata, Dryad). 
The use of such repositories is encouraged by various publishers (such as APS and \textit{Science}). 

The majority of the conference participants agreed that increased sharing of research details would improve reproducibility and replicability. 
Some observed that confusion and disagreements about the extent and specific means of data sharing were obstacles. 
The duty to share is also not clear to everyone in the community. 
A common statement in the journals is that data will be available upon a “reasonable” request. 
This begs the question: what would be considered a “reasonable” request? 
And what would make a request “unreasonable”? 
The ambiguity of this phrase leaves too much room for some authors to justify a refusal to share data. 

At the same time, the wider use of data repositories is hindered by various social factors. 
Many researchers are unaware of the technical resources available to them for data sharing, or have not adopted a group policy for archiving and publicly sharing data. 
Others are concerned about getting “scooped” if they share more than the minimal amount of information~\cite{Laine17, Gomes22}.
Some authors feel reluctant to share their data files or computer code and scripts because they do not consider the files “tidy enough” to be read and understood by others. 
As a result, open sharing is perceived as additional overhead for researchers because it requires extra effort to organize the data into a “useful” repository. 

Various stakeholders also recognize the need to balance the principles of open science against the considerations of individual privacy, commercial interests, and national security.  
Some authors cite IP restrictions as a reason not to share data and samples. 
Such arguments would seem unjustified in cases where the authors did not disclose the existence of such limitations in their paper (e.g., in the ``Competing Interests'' section). 
But in practice, many editors and scientific colleagues are unsure how to respond when the authors voice IP concerns. 
We note that United States copyright law protects creative works, such as manuscripts, posters, and computer code, but not data. 
Jurisdictions outside of the United States have different intellectual property laws~\cite{Carroll15, Labastida20}.

%%%%%%%%%%%%%%%%%%%%%%%%%%%%%%%%%%%%%%%%%%%%%%%%%%%%%%%%%%%%%
\subsection{Publishing policies}
%%%%%%%%%%%%%%%%%%%%%%%%%%%%%%%%%%%%%%%%%%%%%
Publishers have the opportunity and, in light of their role in disseminating results, likely also the responsibility to verify and enhance the reproducibility and replicability of the claims they publish. 
While they could be doing more today, there are positive indicators for reproducibility and replicability in some of the policy changes being rolled out by the publishing organizations. 
For example, at \textit{Science} magazine, a policy was established in 2018 that requires data published in a paper to be made available in tabular form. 
However, there is still no policy in place to make the original raw data or fuller data available. 
\textit{Nature} currently has a policy that calls for the sharing of a \textit{“minimum dataset that is necessary to interpret, verify and extend the research in the article, transparent to readers”} (Nature Portfolio, Reporting standards and availability of data, materials, code and protocols). 
However, such a standard is open to wide interpretation and, in practice, is often ignored by authors and editors. 
While there are some field-specific policies regarding which methods, data types, and so on must be shared, none of these policies are presently specific to CMP. 
APS Publishing recently revised its initial data availability policy for some journals, requiring authors to explain where the data can be found, though it does not mandate data sharing~\cite{GDA}. 
IOP Publishing encourages authors to cite any data referred to in the article (including the authors’ own data) in the reference list. 
However, they also state that \textit{“[t]he decision to accept an article for publication will not be affected by whether or not authors share their research data publicly.”} 

Journals have taken steps that focus on broader cultural and ethical aspects, which can indirectly benefit reproducibility and replicability. 
Most major journals where physicists publish are members of the Committee on Publication Ethics (COPE), which provides guidelines and some oversight for specific cases and publication disputes. 
At the same time, COPE can be viewed as an industry self-policing entity and may therefore be prone to bias in protecting publishers. 
Several publishers, including APS, American Association for Advancement of Science (AAAS), and Springer-Nature, have signed the Declaration on Research Assessment (DORA), which contains a number of progressive ideas that should aid reproducibility~\cite{DORA}. 
As with many policy initiatives, there is a delay between signing on and taking concrete steps to implement the principles in everyday decision-making. 
However, the hope is that the first changes will become apparent in the not-too-distant future.  
To bring this moment closer, publishers should ensure that research ethics policies are easily accessible on the journal website and follow a comprehensive, practical, actionable, and list-like format.

Despite these steps, journals are still perceived as insufficiently responsive to integrity issues. 
Many publishers take the stance that they merely provide means of communication about science and are not in the business of verifying reproducibility or replicability. 
Some of the journals in their descriptions of the referee roles use language that appears to actually discourage extensive error-checking~\cite{Subbaraman23, PNAS}. 
While journals have published several negative replication studies in CMP, a practice that should be praised~\cite{Yu21, Kayyalha20, Valentini21, Valentini22}, these are very rare and often appear in a lower-impact product of the same publisher. 
This type of publication often confuses referees, who are accustomed to evaluating claims of novelty and breakthroughs and are not provided sufficient context to properly assess the value of a replication study.

In the initial peer review, most journals do not ask referees to assess reproducibility. 
Referees are also not provided with sufficient information to verify the reliability of various claims. 
Indeed, sharing data is still not a common practice, so a referee would need to make a data request through the editors, which is considered unusual. 
An author-driven practice that is becoming increasingly common is the addition of Supplemental Material  to the manuscript. 
This can be a useful source of information for replication; however, there is no standard for what those documents should contain, and as a result, Supplemental Material  vary drastically from paper to paper. 
In reality, referees consider these materials at their discretion, and their evaluations are likely inconsistent. 
At the same time, the confidentiality of the peer review discussions means that readers are unaware whether all reviewers deemed the results likely to be replicable and reproducible, and based on what evidence.

The problems continue past publication. 
Several authors of this report raised concerns that when they attempted to submit a commentary on replicability challenges in published work, they found journals slow to consider the submission, investigate more serious accusations, or issue a correction or retraction. 
Some of this delay is understandable and is due to the complex and delicate nature of the situation, as well as the need to coordinate with the institution where the work was performed.
However, the general lack of a clear, established procedure for assessing these concerns leads to failure to act even in egregious situations. 
Journals are often impeded by the perceived likelihood of legal threats from the original authors. 
Even though these are difficult to follow through on, at least in the United States, they can have a chilling effect on the journal or the complainants.

After publication, obtaining additional data is even harder without the leverage offered by the pending acceptance decision. 
Journal editors rarely facilitate post-publication data requests from peers, even when their policies expressly require authors to cooperate. 
Even when data requesters ask the editors to do no more than forward the request to the original authors, editors are hesitant to do so, fearing they would be applying undue pressure or getting involved in a dispute.
The authors’ willingness to cooperate with verification efforts is inconsistent. 
It is true that in some cases, granting access to equipment or original samples would take significant time and resources on the part of the authors. 
It is harder to understand some authors’ reluctance to facilitate access to raw data that can be easily placed online or in a private data-sharing repository. 
An editor would seem to be on the moral high ground if they required their authors to cooperate with this kind of data sharing, and even to use retraction as a consequence for noncompliance, a policy employed by \textit{Proceedings of the National Academy of Sciences} (PNAS).

%%%%%%%%%%%%%%%%%%%%%%%%%%%%%%%%%%%%%%%%%%%%%%%%%%%%%%%%%%%%%
\subsection{Research institution and government policies}
%%%%%%%%%%%%%%%%%%%%%%%%%%%%%%%%%%%%%%%%%%%%%
Primary research institutions have considerable work to do to ensure the reproducibility and replicability of research conducted within their walls, even just to comply with existing government regulations. 
Currently, institutions often act as impediments to reproducibility and replicability efforts, both by failing to resolve disputes over potential research integrity violations promptly and transparently, and by taking few measures to foster and encourage an internal culture that values reproducible work.

Institutions are establishing data policies, which, in principle, stipulate where data should be stored. 
Some of these require little more than data sharing upon request. 
However, even these measures are not enforced and are left up to individual researchers’ discretion. 
Even though funding agencies have regulations enabling access to data and frequently require data management plans, researchers and institutions do not always follow through on these, and agencies find them difficult to enforce.

Some institutions have actually introduced regressive policies. 
For instance, the University of Pittsburgh states, in a policy introduced in 2023, that \textit{“[...]Research Records shall be available only to those who need such access and to the minimum amount necessary”}~\cite{UPitt23}. 
Other organizations and entire nations, such as Denmark and the Netherlands, promote full data sharing, though this is not always followed by institutions and researchers in those countries.

There is a growing expectation among government funders for accountability, transparency, and durable information stewardship, with data and code sharing seen as key to advancing reproducibility and replicability.
As Musen \textit{et al.} (2022) observe, \textit{“National funding agencies increasingly view the results of the research that they support—including the data—as a public good, and they view the availability of FAIR data as the means to deliver to taxpayers the benefit that they have paid for.”}~\cite{Musen22}. 

Governments around the world generally acknowledge that a problem of scientific reproducibility and replicability exists. 
Inevitably, in scarce funding regimes, they do not apply their full capacity to formulating the clearest, most unified policy for replication. 
Nor do they do enough when it comes to enforcement. 
In the United States, the Office of Science and Technology Policy (OSTP) designated 2023 the “Year of Open Science” and released a number of policies designed to improve reproducibility across the disciplines~\cite{FSBHA}. 
Aligned with the OSTP action, the United States National Science Foundation has issued a Dear Colleague Letter (DCL) “Reproducibility and Replicability in Science” (nsf23018). 
At that time, NSF already had a long-standing data sharing requirement written into its Proposal \& Award Policies \& Procedures Guide (PAPPG)~\cite{NSF} dating back to 2015 when it introduced the Public Access Plan~\cite{NSF-2}. 

Across the board, enforcement is lagging very far behind the stated government goals. 
For instance, the NSF Office of the Inspector General (OIG) received only 21 allegations of fabrication or falsification across all programs in FY 2023 and opened only 12 investigations. 
In the Netherlands, similarly low numbers of research misconduct investigations pass through the national research integrity body (LOWI in Dutch), and most findings are classified as “ungrounded”.  
This is likely a small percentage of situations based on open online sources such as Retraction Watch Database~\cite{RWD04}.

Research misconduct, which is a big driver of the replication crisis, has been a particular focus of the APS during the past 20 years~\cite{Houle23, Kirby04}, and the APS has established ethical guidelines concerning publication and treatment of colleagues~\cite{APSEG}. 
While ethics training has become more common in graduate programs~\cite{Houle23}, the prevalence of research misconduct remains relatively high and has not declined significantly during the past two decades~\cite{Baker16}. 

When claims of research misconduct are raised, the ensuing institutional investigations are often marred by low-quality reviews, conflicts of interest, delay tactics, and the obfuscation of responsibilities. 
Ambiguous language and nontransparent procedures have contributed to ineffective investigations of unreliable work in some cases. 
We note that even when clear government or professional society policy exists, such as a code of conduct, institutions find ways to circumvent it or effectively disregard it.

%%%%%%%%%%%%%%%%%%%%%%%%%%%%%%%%%%%%%%%%%%%%%%%%%%%%%%%%%%%%%
\section{Detailed recommendations of the report}
%%%%%%%%%%%%%%%%%%%%%%%%%%%%%%%%%%%%%%%%%%%%%%%%%%%%%%%%%%%%%
\subsection{Best practices for condensed matter researchers}
%%%%%%%%%%%%%%%%%%%%%%%%%%%%%%%%%%%%%%%%%%%%%
Below, we list recommended best practices for CMP scientists to enhance the reproducibility and replicability of their work. 
The APS Ethics Guidelines for research results include some of these recommendations~\cite{APSEG}, but the list here goes beyond them. 
The recommendations, although tailored for the CMP community, should be applicable more broadly.

\textit{When reporting experiments, include the full ranges of experimental parameters}. 
Theoretical studies should similarly report the full parameter ranges explored, which should be sufficient to enable an independent assessment of how robust or exotic the theory's approximations and assumptions are. 
The practice of exploring the full range of parameter space safeguards against real-time confirmation bias, which can result in reporting exciting conclusions based on a narrow range of data. 
When preparing a manuscript, disclose full parameter ranges studied, and discuss any limitations or omissions from the full possible ranges permitted by the experimental apparatus.

\textit{All authors of a manuscript should have access to all primary data and analysis files/software before they are asked to consent to publication.} 
At a minimum, all authors, no matter their expertise, should have the opportunity to review the data and reproduce the analysis of a paper before the paper is published. 
Nonexperts should actively seek to understand the parts of the paper written by others. 
For example, a theorist can seek a deeper understanding of the data behind the experimental part, including how the data were chosen for the figures. 
An experimentalist can try to understand the work done by the sample grower. 
Data access should include the ability to review laboratory notebooks and internal presentations that facilitate understanding of the work of the co-authors. 
Furthermore, for multi-author papers, we recommend that every element of the paper be reviewed by at least two authors before public dissemination. 
This is especially relevant when the project crosses the boundaries of traditional silos of expertise, such that not every coauthor is an expert in all aspects of the study.

\textit{The manuscript, its Appendixes, and Supplemental Material should contain enough information for replication.} 
Experimental and theoretical works may involve complex instrumentation or analytical methods, but authors should ensure they provide sufficient information so that a motivated reader with similar training can reproduce their results. 
Makes and models of instruments used to produce samples and collect data should be included, as should information about the specific software used (including exact versions of all external libraries in case projects are not actively maintained).

\textit{Authors should release as much primary data as possible and practical with a paper.} 
We encourage authors to share all primary data acquired in the experiments presented in the paper, including metadata. 
In many cases, sharing full data is straightforward. 
If a technical obstacle prevents data release, the authors should explain why (for instance, because the volume exceeds repository limits). 
In such situations, authors should provide sufficient data to facilitate replication efforts.  
Authors should use citable repositories to share their data.

\textit{Strive to achieve FAIR (Findable, Accessible, Interoperable, and Reusable) Data Standard, but, if not possible, share materials in the form you have them.} 
Since 2016, the FAIR Principles have offered a framework for data sharing across disciplines~\cite{Wilkinson16}. 
FAIR research outputs are more likely to be reproducible. 
At the same time, achieving full FAIR compliance may be too high a standard for a community like ours, which does not do this habitually. 
Requiring high standards for data sharing, no matter the circumstances, could create a barrier to any data sharing. 
Nonetheless, where elements of the FAIR standard can realistically be met without imposing a significant burden, this should be done.
 
\textit{A manuscript should include all reasonable presentations of the data.} 
The way authors choose to plot their data (i.e., on a linear or logarithmic scale, or in a colorscale or line graph) can be used to favor one interpretation over another. 
We recommend that authors provide all reasonable alternative presentations of the data (for example, in the Supplemental Material). 
At the very least, basic, unprocessed, or minimally processed data plotting should be included. 
This effort can also be supplemented by sharing computational notebooks/scripts and primary data (e.g., Jupyter notebooks), which can be replotted by a reader.  

\textit{Authors should specify full data processing steps from original raw data to published figures.} 
We recommend that, for each paper, the authors provide (via a public repository or as Supplemental Material) computer scripts that process the figures and produce them from the original raw data. 
Alternatively, authors can specify the software (and version) used to produce the figures and include the project file that can be used with that software.  

\textit{A manuscript should summarize all plausible interpretations of results.} 
Selective interpretation of results is a significant problem in CMP, leading to the propagation of novel narratives when more straightforward explanations are likely~\cite{Mazin22}. 
We recommend that every paper include a discussion of all known plausible alternative explanations for the data. 
And, as much as possible, the manuscript should discuss how the study provides evidence for and against each of them. 
While the authors may think that a more confidently written paper would fare better in peer review, the growing skepticism about overly simplistic presentations of results and rising awareness of replicability issues are likely to shift expectations toward a more balanced presentation of results.

Similarly, theory papers that are “inspired” by an experimental result, or that offer an explanation of an experiment, should contain a discussion, even if brief, of other possible explanations of that experiment. 
Authors should also always include a critical discussion of the assumptions behind their model, particularly if they are using it to justify extraordinary conclusions. 

Finally, authors are encouraged to separate the presentation of data from their discussion of the methodology used to process and analyze it, as well as from claims the data are meant to support. 
This principle is well understood in the field, but in practice, pressure to present novel results often leads to the mixing of data presentation and interpretation, making it difficult for readers to determine what is actual data and what is effectively little more than an illustration of the authors’ preferred interpretation.

\textit{Information about the number of samples tested should be reported.}  
When multiple samples or devices are examined, authors should report the number of samples or devices tested. 
These statements should include details, such as the total volume of data collected and the duration of the study. 
We recommend that the authors also discuss how their second- and third-best datasets appear, to help readers assess the degree of reproducibility within the study. 
We encourage the authors to highlight nonconfirmatory samples or data and openly discuss possible explanations for them.
    
\textit{Research groups should have a stated policy for eventually releasing in-house developed code.} 
Some research groups invest considerable effort into developing in-house source code for numerical simulations or calculations. 
This source code constitutes a significant competitive advantage for the group, and it may not be fair to ask the group to release their code publicly at the same time as the manuscript that presents their numerical results. 
In such cases, we recommend developing a delayed or embargoed code release policy for a public codebase and declaring it in the manuscript. 
For example, authors can upload their code to a private codebase (potentially maintained by the journal itself), which is made public after a specified period. 
Whatever the approach, each paper should state the steps the authors are willing and able to take toward the reproducibility and replicability of their numerical results.

\textit{Subcommunities of condensed matter should develop subfield-specific minimum reporting standards.} 
Certain scientific disciplines and subfields have developed checklists of minimum reporting standards to ensure reproducibility of published work. 
For example, for biomedical randomized controlled trials, there are the Consolidated Standards of Reporting Trials~\cite{Moher01}. 
Checklists have been developed for other fields, including statistics in biological sciences~\cite{NPC}, solar cell research~\cite{NRSCRC}, battery research~\cite{Mistry21}, and laser research~\cite{NRLRC}. 
We encourage subfields of condensed matter to form working groups that devise similar principles, checklists, or handbooks~\cite{Zwolak24} for reporting within their subfield. 

\textit{Authors should promptly retract or correct their own prior work if they discover errors post publication.} 
As laid out in the APS Ethics Guidelines~\cite{APSEG}, it is the obligation of each author to produce prompt retractions or corrections to errors in published work. 
Such behavior should be praised rather than stigmatized, provided the reasons for correction or retraction are accurately communicated.

%%%%%%%%%%%%%%%%%%%%%%%%%%%%%%%%%%%%%%%%%%%%%%%%%%%%%%%%%%%%%
\subsection{Changes to scientific publishing}
%%%%%%%%%%%%%%%%%%%%%%%%%%%%%%%%%%%%%%%%%%%%%
\textit{Refocus on scientific validity rather than subjective appeal criteria for publication.} 
While significance and novelty are important aspects of publication value, the prioritization of subjective criteria creates, among other things, pressure to produce sensational data presentations and flashy stories, and to mix data and interpretation. 
The long-term utility of a result may derive from knowledge that is not available at the time of publication. 
We recommend that editors prioritize objective criteria, such as clear problem statements, high-quality data presentation, and above all, accuracy. 
These requirements should be emphasized to referees and made known to the authors of submitted manuscripts.

\textit{Require availability of complete data and scripts in citable repositories.} 
Instead of “data available upon reasonable request” statements, which still make their way into published papers, sometimes contrary to existing journal policy, we recommend mandating full data availability from the entire study in a citable repository as well as sharing of methods, and analysis scripts needed to reproduce processed data and images, and to arrive at the same conclusions as in the paper. 
Exceptions can be made when privacy, commercial interests, or national security concerns are stated upfront. 
To reduce the chance of authors unwillingly participating in nonreproducible data analysis, all authors should be required to certify that they were provided access to the data prior to publication.

\textit{Formulate clear statements of acceptance criteria for replication efforts.} 
Many journals are reluctant to publish replication efforts, hindering the correction of mistakes. 
To mitigate publication bias, we recommend that publishers establish policies to facilitate the publication of replication research, including the creation of a new article type or section to promote such work. 
Such papers should be linked to the article they are commenting on and be published in the same journal as the original claim, especially in situations where earlier conclusions are found not to be replicable. 
Editors should be encouraged to take pride in soliciting and publishing replication efforts, and should consider promoting the results, even when they are negative.

\textit{Create review requirements that are specific to Supplemental Material.} 
The science presented in the online Supplemental Material  must be adequately reviewed and scrutinized. 
We recommend providing referees with clear statements outlining the review criteria for such material, including guidance on evaluating various types of data. 
A complementary recommendation is to create pathways for publishing such material as primary publications in their own right, such as the “Physical Review Letters + Physical Review B” joint publication or follow-up submissions. 
Regardless of the approach, promoting a more detailed presentation of each study is advisable.

\textit{Develop specific guidelines for assessing the reproducibility and replicability of submissions.} 
Guidelines to review reproducibility and replicability are often lacking or insufficient. 
We recommend developing guidelines for reviewing the availability of scripts, Supplemental Material, and data. 
These guidelines should ensure that enough information is provided to facilitate attempts to reproduce or replicate the published work. 
Specifically, we encourage journals to include some or all of the recommended best practices listed above in a checklist given to referees for assessing reproducibility and replicability.

\textit{Consider adopting an open review process.} 
Traditional peer review is opaque and may not evaluate reproducibility and replicability. 
We recommend considering an open review process, including: 
\begin{enumerate}[label=(\arabic*), topsep=-3pt, itemsep=-3pt]
    \item Open participation review---anyone can referee before or after the paper is published (post-publication review option).
    \item Open reports review---referee reports are publicly available during review or after publication;
    \item Open identities review---referees may opt to sign their reports.
\end{enumerate}

Adopting one or more of these options could go a long way in addressing the problem of a lack of expertise or incentives to review the reproducibility of the work by enlarging the referee pool through open participation and giving credit via open identity review. 
A concern with the “open identities” review may be the unintended creation of quid-pro-quo reviews. 
However, we do not see an inherent bias toward such unethical practices compared to existing anonymous review processes, in which referees can still reveal their identities to authors if they choose to. 
A more pressing concern is the inhibition of candid reviews by junior scientists, which should be addressed by making identity disclosure optional.

\textit{Draft and publicize more transparent retraction and correction policies and procedures}. 
Stigma around retraction discourages correcting papers. 
This trend is reinforced by practices at editorial offices that process retractions without sufficient transparency regarding the procedural steps being undertaken~\cite{Resnik13}. 
We recommend specific, clear retraction policies and procedures, including timelines for decision-making, the bases for decisions, and the types of situations in which expressions of concern, rather than retractions, are issued.
Publishers should provide avenues for efficiently communicating community feedback on published papers. 
The overly complicated process for submitting \textit{Comments}, \textit{Matters Arising}, and other formal, written responses should be simplified, and their publication expedited. 
It should also be considered best practice to include all co-authors in communications that question the reproducibility or replicability, corrections, or retractions. 
Should, at any point in time, any of the authors no longer stand by the paper’s conclusions, these authors should have an easy way of registering this change in their position and the basis for it.

%%%%%%%%%%%%%%%%%%%%%%%%%%%%%%%%%%%%%%%%%%%%%%%%%%%%%%%%%%%%%
\subsection{Recommendations for universities and research laboratories}
%%%%%%%%%%%%%%%%%%%%%%%%%%%%%%%%%%%%%%%%%%%%%
\textit{Consider not only publications, but also datasets and code as research outputs.}  
Ask for datasets and code made publicly available to be listed on CVs, in PhD theses, and encourage their submission to funding agencies in grant reports. 
Strive to organize them using FAIR principles as a target and help investigators achieve this standard. 

\textit{Recognize reproducibility and replicability practices in promotion cases.} 
Promotion is based in part on scholarly achievement and in part on academic leadership. 
Making one’s work more replicable and encouraging reproducibility and replicability in one’s field can fall through the cracks in established promotion criteria. 
We recommend adjusting the criteria so that this type of activity can be formally recognized as credit to the investigator.

\textit{De-emphasize publication indices in evaluating the productivity of individual scientists.} 
Publication metrics, such as citations, can help identify relevant research. 
However, when used to assess the importance of individual scientists’ research, they lead to perverse publication incentives that distort the scientific process. 
These considerations have led to the establishment of DORA, which has been signed by, among others, the APS, the AAAS, and Springer Nature. 
Included in DORA are the recommendations that we strongly support:
\begin{enumerate}[label=(\arabic*), topsep=-3pt, itemsep=-3pt]
    \item When involved in committees making decisions about funding, hiring, tenure, or promotion, make assessments based on scientific content rather than publication metrics.
    \item Challenge research assessment practices that rely inappropriately on journal impact factors and promote and teach best practice that focuses on the value and influence of specific research outputs.
     \vspace{5pt}
\end{enumerate}

\textit{Follow government-mandated procedures for investigating misconduct, and inquire into the full factual basis of each report.} 
Universities, national labs, and commercial labs are not sufficiently incentivized to investigate reports of scientific misconduct. 
The reasons are varied but include the belief that misconduct tarnishes reputations and that investigations are time-consuming and expensive. 
The institutions where research was conducted are fundamentally conflicted in carrying out the subsequent investigations. 
However, often these institutions are the only ones in possession of the primary materials, communications between the authors, original equipment, and other materials essential to the investigation.  
A thorough investigation of potential scientific misconduct or any erroneous presentation of research is crucial to maintaining the integrity of the scientific process.  
Institutions should face external consequences for failing to conduct thorough investigations. 
In anticipation that this pressure will only increase, they should consider the full factual basis for each report of suspected misconduct and establish a transparent, well-publicized process for investigations. 
Institutions should communicate any concerns about research to the journals and funding agencies in a timely and efficient manner.

Institutional processes must be carried out rigorously and with minimal local conflict of interest, or perception thereof. 
Wherever possible, inquiries should be conducted by an independently appointed inquiry committee that ideally includes external peer experts. 
Where misconduct is found, penalties must be enforced and fairly applied. 

\textit{Develop a clear definition of unreliable science beyond the definition of misconduct.} 
Reducing the production of unreliable science at the institutional level requires a clear understanding of what constitutes unreliable science. 
Often, the standard fabrication-falsification-plagiarism definition developed for cases of misconduct is the only available standard. 
This standard does not include many practical examples of unreliable research, such as technical errors or interpretive errors, for which the institution is ultimately responsible but which individual affiliated authors may not be motivated to correct. 
It is common for an institution to find no misconduct in an investigation and therefore take no further action, allowing unreliable research to stand and propagate. 
The relevant research integrity officers at the institution should be actively aware of cases of unreliable research and be familiar with methods for characterizing and identifying them. 
In the absence of proof of misconduct, requiring correction of unreliable research that has led to a complaint can also help resolve disputes in a timely manner. 
This is already required by the APS ethics code, but is not uniformly enforced by the institutions.

\textit{Actively protect whistleblowers.} 
Scientists expressing concerns or questions about reproducibility, replicability, or data access should be granted whistleblower protections, whether or not they self-identify as whistleblowers. 
Protecting those who challenge or report questionable science is difficult but essential for maintaining research integrity. 
The most vulnerable whistleblowers are often graduate students, postdoctoral researchers, and junior collaborators. 
There are many reports of such individuals feeling intimidated or manipulated into going along with claims they have reason to doubt. 
Institutions should include an independent scientific integrity ombuds office separate from the research integrity officers that can advise junior researchers on their options, including how to report research integrity concerns safely. 
Keep a list of nonprofit organizations that can provide help, including legal defense for whistleblowers.

%%%%%%%%%%%%%%%%%%%%%%%%%%%%%%%%%%%%%%%%%%%%%%%%%%%%%%%%%%%%%
\subsection{Measures for government and scientific funding organizations}
%%%%%%%%%%%%%%%%%%%%%%%%%%%%%%%%%%%%%%%%%%%%%
\textit{Consider funding work that explicitly aims to replicate prior results.} 
The cost-savings potential of efforts to replicate results that can enable the timely correction of unreliable research claims is high. 
Yet, attempts to reproduce results, even those of perceived exceptional importance, are costly and time-consuming when considered independent of their potential. 
In the current CMP culture, such studies are often deemed “not worth it” because the potential professional rewards are low. 
Funding organizations can help change this culture by explicitly funding work to verify or replicate high-profile claims, particularly those on which the funding organization relies when deciding which follow-up research to support.
    
Funders can also consider providing supplemental funding to work that is incidental to a funded project but which promotes reproducibility and replicability. 
For example, funders could provide supplemental funding to support researchers in sharing materials, samples, code, or information about the experiments. 
For instance, a crystal grower may need additional support to provide samples to groups interested in replication.   

Beyond support for direct replications, it is important to recognize work that yields negative results or results that may undermine earlier claims. 
Funding organizations should make a dedicated effort to reward such research and thereby remove the stigma that the academic community sometimes associates with it.

\textit{Require grant proposals to address reproducibility and replicability in their planning.} 
At present, there is no uniform mandate from funding institutions to share data, methods, code, and related materials associated with published work. 
We encourage funding bodies to establish clear criteria for addressing reproducibility and replicability issues in funded work. 
Specifically, we encourage funders to review the list of \textit{Best Practices for Scientists} we provide here and to incorporate some or all of these requirements as conditions for funding. 
    
Data sharing and management activity should be reviewed through annual reports, which could include a listing of data repositories produced during the reporting period. 
The agencies can keep track of these repositories by requesting that they cite the grants that supported their creation.

\textit{Evaluate and reward reproducibility and replicability of research.} 
Grant reviewers should be provided with clear guidelines on how to evaluate reproducibility and replicability criteria as elements of the merit criteria and the scientific management of the project. 
Proposals that specify steps to facilitate the reproducibility and replicability of the published work, for example, by committing to open sharing of code or materials, should be encouraged. 
Reproducibility and replicability should not be optional; namely, it should not be possible to compensate for a failure to commit to these steps by promising to undertake other kinds of work, such as broader-impact activities.

\textit{Develop policies requiring grant recipient organizations to report allegations of scientific misconduct to the funding agency.} 
When the investigation is handled by the organization, a report of any investigation or inquiry must be submitted to the funding agency, clearly outlining the path to the final conclusion. 
The agencies should consider requiring that at least one external scientist with expertise in the questioned area serve on any investigative committee. 
At present, the organizations submit only the findings of misconduct, leaving much of the investigative process to their discretion and limiting oversight and transparency.

%%%%%%%%%%%%%%%%%%%%%%%%%%%%%%%%%%%%%%%%%%%%%%%%%%%%%%%%%%%%%
\section{Outlook}
%%%%%%%%%%%%%%%%%%%%%%%%%%%%%%%%%%%%%%%%%%%%%%%%%%%%%%%%%%%%%
We find many indications that CMP is not immune to the well-recognized replication crisis in the social and biomedical sciences.
Adopting research practices that promote reproducibility and replicability not only helps preserve the integrity of scientific results but also offers additional benefits, including new synergies, more efficient use of public resources, and long-term preservation of research outputs.

\textit{What do we need to improve reproducibility?} 
The most straightforward way to enhance reproducibility across the field is to adopt a proactive approach to sharing primary materials, including data, code, protocols, and, where possible, samples. 
The goal is to ensure that the entire analysis process can be tracked from the original data to the figures in the paper by a third party.
The ideal situation is when all materials necessary to reproduce the findings are publicly available prior to any peer review, so that reviewers can verify the steps followed, even if they fall short of a comprehensive replication. 

The technical means for sharing primary research records exist due to advances and the ready availability of high-capacity online storage. 
The next step should be to promote sharing these materials at all levels: individual labs, institutions, publishers, professional societies, and government agencies. 
Policies should be straightforward to follow and enforce, and the enforcement should not be optional. 
At the same time, career advancement and funding decisions should reward reproducibility.

\textit{What do we need to improve replicability?} 
The first condition for replicability is, of course, reproducibility.
It would not make much sense to try to replicate a study if the study itself cannot be reproduced from its own data. 
However, the key requirement for improving replicability efforts is the willingness of funding agencies to support studies aimed solely at replicating previous findings, and of journals to publish such works.

We are hopeful that these measures will lead to a change in the culture and practice of science. 
We suggest developing best practices and referee checklists specific to the most active subfields in CMP and quantum science. 
We expect periodic assessments of the state of replicability in the larger field and in at least the most active subfields to be conducted, using methods developed in other fields but adapted to the specific needs of CMP.

%%%%%%%%%%%%%%%%%%%%%%%%%%%%%%%%%%%%%%%%%%%%%%%%%%%%%%%%%%%%%
\begin{acknowledgments}
Work on this report and the conference was supported by the United States National Science Foundation under Grant No. DMR-2326983. 
We acknowledge support from the Julian Schwinger Foundation for Physics Research and from the Pittsburgh Quantum Institute during the conference. 
We thank E. Reich for the critical reading of the manuscript. 
The opinions, recommendations, findings, and conclusions presented in this report do not necessarily reflect the views or policies of the National Institute of Standards and Technology or the United States Government.
\end{acknowledgments}
%%%%%%%%%%%%%%%%%%%%%%%%%%%%%%%%%%%%%%%%%%%%%%%%%%%%%%%%%%%%%

%%%%%%%%%%%%%%%%%%%%%%%%%%%%%%%%%%%%%%%%%%%%%%%%%%%%%%%%%%%%%

%%%%%%%%%%%%%%%%%%%%%%%%%%%%%%%%%%%%%%%%%%%%%%%%%%%%%%%%%%%%%

%%%%%%%%%%%%%%%%%%%%%%%%%%%%%%%%%%%%%%%%%%%%%%%%%%%%%%%%%%%%%
\end{document}